\documentclass[aps,prc,superscriptaddress,twoside,twocolumn,nofootinbib,%
showpacs,floatfix]{revtex4}

\usepackage{amssymb}
\usepackage{graphicx,bm}

\begin{document}

\title{\boldmath On the sign of the $\pi\rho\omega$ coupling constant}

\author{K. Nakayama}%

\affiliation{Department of Physics and Astronomy, University of Georgia,
Athens, GA 30602, USA}
\affiliation{Institut f\"ur Kernphysik, Forschungszentrum
J\"ulich, D-52425 J\"ulich, Germany}

\author{Yongseok Oh}%

\affiliation{Department of Physics and Astronomy, University of Georgia,
Athens, GA 30602, USA}

\author{J. Haidenbauer}%

\affiliation{Institut f\"ur Kernphysik, Forschungszentrum
J\"ulich, D-52425 J\"ulich, Germany}

\author{T.-S. H. Lee}%

\affiliation{Physics Division, Argonne National Laboratory,
Argonne, IL 60439, USA}

\date{}

\begin{abstract}
It is shown that the relative sign between the $NN\omega$ and $\pi\rho\omega$
coupling constants can be determined most sensitively from $\omega$ production 
processes in $NN$ collisions.
Recent data on these reactions clearly favor the sign of the $\pi\rho\omega$
coupling constant which is opposite to that inferred from studies of the
photoproduction reaction in combination with the vector meson dominance
assumption and used by many authors.
Implication of this finding in the description of other reactions is discussed.
\end{abstract}


\maketitle


A reliable description of hadronic reactions involving the production of 
$\pi$, $\rho$ or $\omega$ mesons based on meson and baryon degrees-of-freedom
usually requires consideration of the $\pi\rho\omega$ axial anomaly coupling,
due to its relatively strong coupling strength which makes reaction mechanisms 
involving this vertex quite important.
The effective Lagrangian density corresponding to this coupling can be
written as
\begin{equation}
{\cal L}_{\pi\rho\omega} = \frac{g_{\pi\rho\omega}^{}}{\sqrt{m_\rho^{}
m_\omega^{}}}
\varepsilon_{\alpha\beta\mu\nu}
\partial^\alpha\vec\rho^{\,\beta} \cdot \partial^\mu\vec\pi\omega^\nu \ ,
\label{pirhoomega}
\end{equation}
where $\vec\rho^{\,\beta}$, $\vec\pi$, and $\omega^\nu$ denote the
$\rho$, $\pi$, and $\omega$ meson fields, respectively.
The vector notation refers to the isospin space and
$\varepsilon_{\alpha\beta\mu\nu}$ denotes the Levi-Civita antisymmetric 
tensor, with the convention%
\footnote{Some authors use the convention $\varepsilon_{0123}=+1$
(or, equivalently, $\varepsilon^{0123}=-1$), which is opposite to ours.
In the current discussion, those have been converted to the present
convention. Some other authors do not provide information on their phase
conventions explicitly, which makes it difficult to find out the actual
relative sign associated with their $\pi\rho\omega$ couplings.}
$\varepsilon_{0123}=-1$. 
$m_\rho$ and $m_\omega$ denote the $\rho$ and $\omega$ mass, respectively,
and $g_{\pi\rho\omega}^{}$ stands for the corresponding (dimensionless)
coupling constant.

A direct experimental access to the coupling constant of this anomalous 
vertex, $g_{\pi\rho\omega}$, is not possible because the decay 
$\omega \to \rho \pi$ is energetically forbidden.
Moreover, in addition to the determination of its magnitude, there is
the problem of fixing its sign.
The present work addresses the latter issue.
In particular, we shall show that the relative sign of this coupling
constant can be inferred most reliably from $\omega$-meson production
processes in nucleon-nucleon ($NN$) collisions.

In many cases, the coupling constant $g_{\pi\rho\omega}$ is extracted in a
purely phenomenological way, e.g., from the radiative decay
$\omega \to \pi^0 \gamma$, in conjunction with the vector meson dominance
(VMD) assumption \cite{Durso87}.
In this case, the sign of the $\pi\rho\omega$ coupling constant is related
to the associated sign of the $\pi\omega\gamma$ coupling constant which, 
in turn, may be determined, e.g., from the relevant meson photoproduction
processes by examining the interference effects between the relevant
production mechanisms \cite{GM93}. 
Indeed, the J\"ulich \cite{GHHS03} and the Gie{\ss}en \cite{FM98,SLMP04}
groups, in their coupled-channel models of meson-nucleon reactions, use the
sign of $g_{\pi\rho\omega}^{}$ in Eq.~(\ref{pirhoomega}) to be positive,
consistent with the sign of $g_{\pi\omega\gamma}^{}$ employed in 
pion photoproduction analyses \cite{NBL90,GM93,SL96,HNK06} as well as in
the studies of deuteron electromagnetic form factors \cite{GH76}.
The coupled-channel model of the Gie{\ss}en group includes also the
$\gamma N$ channel.
On the other hand, the earlier model of the J\"ulich group \cite{KHKS00}
used a negative value of the coupling constant $g_{\pi\rho\omega}^{}$. 
Also, in Ref.~\cite{BKM99}, the reaction $pp\to pp\pi^0$ near threshold has
been investigated using a negative value for $g_{\pi\rho\omega}^{}$.
The contribution of the production mechanism involving the $\pi\rho\omega$
coupling to this reaction is, however, rather small in the
near-threshold-energy region.
In Ref.~\cite{KKW96}, the coupling constant $g_{\pi\rho\omega}^{}$ 
has been extracted from the study of the vector meson decay processes 
into three pions.
There, a negative sign has been adopted for this coupling although, strictly
speaking, only the relative sign between the coupling constants associated
with the direct ($V \to 3\pi$) and the two-step process
($V \to \pi\rho \to 3\pi$) \cite{GSW62} can be determined $(V=\omega, \phi)$.

\begin{figure*}[t!]\centering
\includegraphics[width=0.42\textwidth,angle=270,clip]{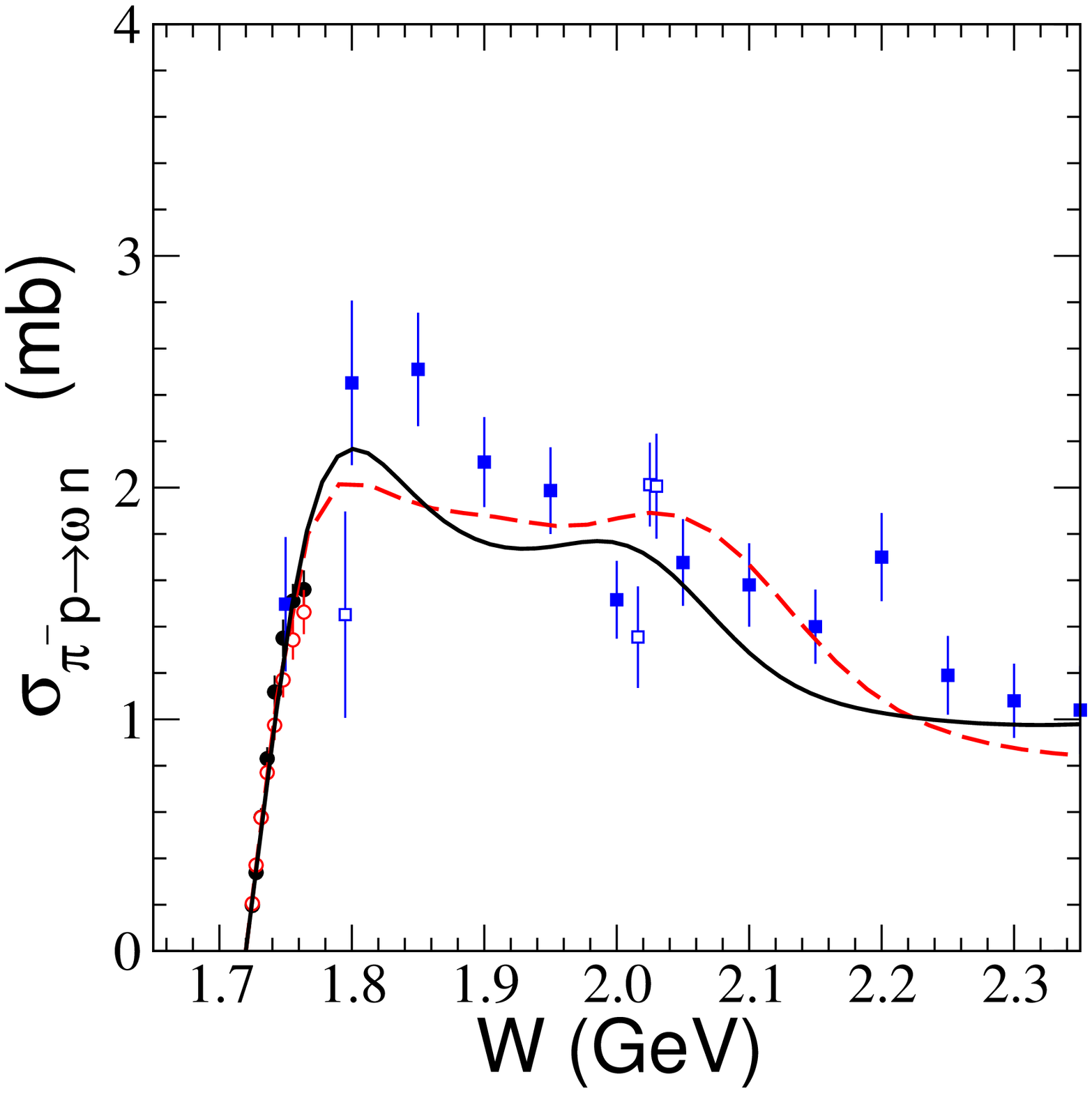}
\hskip -3cm
\includegraphics[width=0.42\textwidth,angle=270,clip]{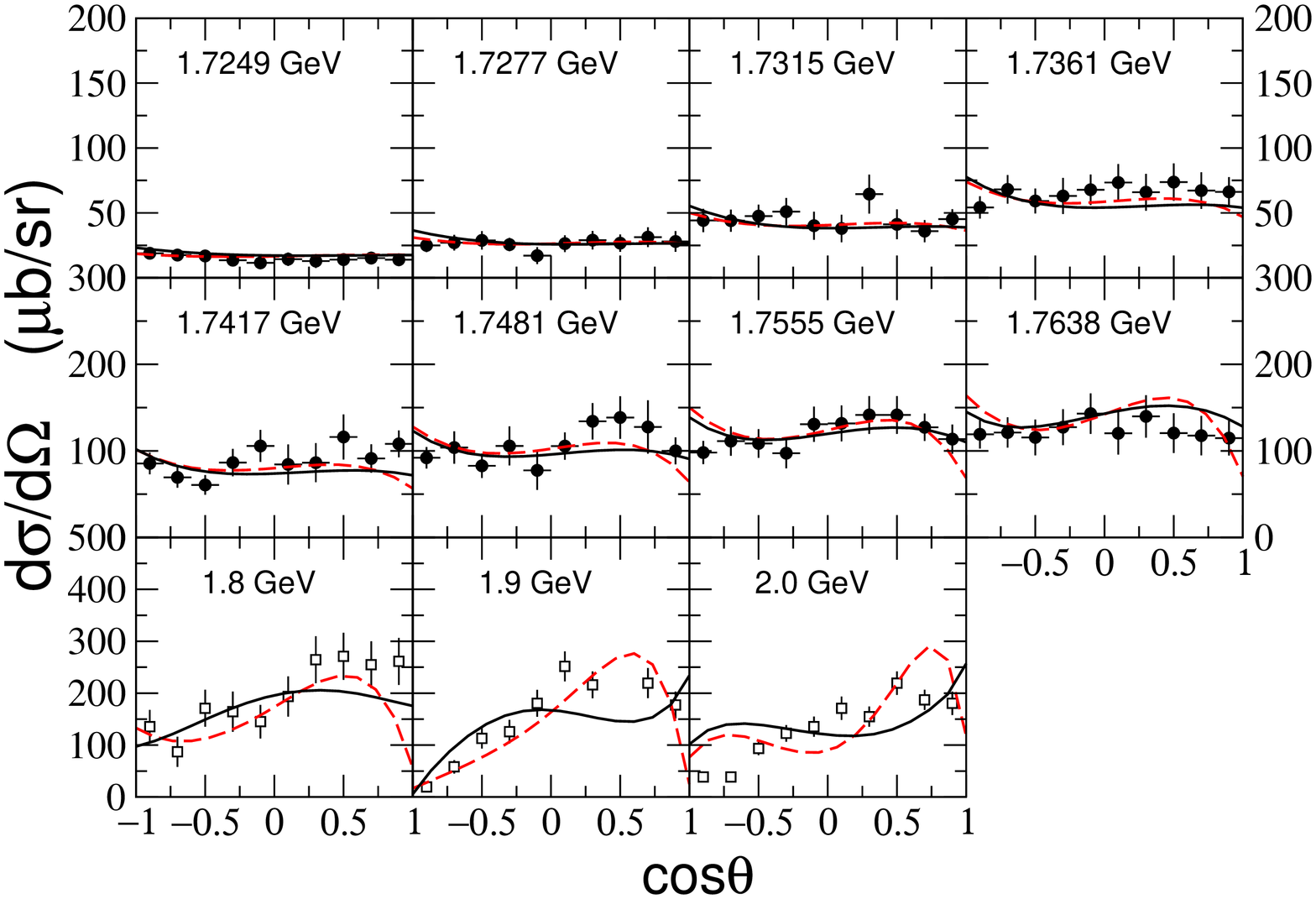}
\caption{\label{fig:piNomegaN}%
Results obtained in the tree-level approximation for $\pi^- p \to \omega n$
\cite{ONL06}, using $g_{\pi\rho\omega}=+11.6$ (dashed curve) and
$g_{\pi\rho\omega}=-11.6$ (solid curve) in Eq.~(\ref{pirhoomega}).
In each case the model parameter values were adjusted to reproduce roughly
the data.
Left panel: total cross section as a function of the total energy $W$. 
Right panel: $\omega$-meson angular distribution in the center-of-mass frame
of the system.
Data are from Ref.~\cite{KCDG79,KBCD76,DADD70}.}
\end{figure*}
%

The difficulty of determining the sign of $g_{\pi\rho\omega}^{}$ is
demonstrated in Fig.~\ref{fig:piNomegaN}, which shows the results of a
tree-level calculation of the total and differential cross sections for the
reaction $\pi^- p \to \omega n$ using both the positive and negative
couplings, $g_{\pi\rho\omega}^{}=\pm 11.6$. 
The model includes nucleonic, mesonic as well as resonance currents.
Details of the calculation will be reported elsewhere \cite{ONL06}.
As can be seen, the results in Fig.~\ref{fig:piNomegaN} are qualitatively
the same for both choices of the sign and illustrate that this reaction is
unsuited for establishing the sign of $g_{\pi\rho\omega}^{}$, at least in 
the energy domain where data exist.
It should be noted that, although both choices of the sign yield similar
results, the corresponding model parameter values, which were adjusted for
each choice of the sign to reproduce roughly the data, are quite different.
It is then clear that the ambiguity in the sign of $g_{\pi\rho\omega}^{}$ 
introduces also uncertainties in the extracted resonance parameters. 
In particular, some resonance coupling constants change their magnitude by a
factor of 1.5 and some couplings also change their signs depending on the
sign choice of $g_{\pi\rho\omega}^{}$.
This evidently shows the importance of knowing the correct sign of
$g_{\pi\rho\omega}^{}$ for investigating the properties of nucleon
resonances.

{}From a more theoretical point of view, the $\pi\rho\omega$ coupling has
been considered within an effective chiral Lagrangian approach for vector
mesons, where the anomalous $\pi\rho\omega$ coupling follows from the
Bardeen-subtracted Wess-Zumino anomalous action.
This approach recovers the result of the low-energy theorem in current
algebra associated with the Adler-Bell-Jackiw anomaly.
The details are reviewed by Mei{\ss}ner in Ref.~\cite{Mei88}.
(See also Ref.~\cite{FKTU85}.)
The corresponding coupling constant $g_{\pi\rho\omega}^{}$ is given by
\begin{equation}
\frac{- g_{\pi\rho\omega}^{}}{\sqrt{m_\rho m_\omega}}
= \frac{3 g^2}{8\pi^2 f_\pi} \ ,
\label{M88:pirhoomega}
\end{equation}
where $f_\pi$ denotes the pion decay constant and $g$ is the universal
gauge coupling constant.
According to this result, the sign of $g_{\pi\rho\omega}^{}$ is manifestly
negative.
However, the specific form of the $\pi\rho\omega$ coupling in
Ref.~\cite{Mei88} assumes a particular realization of VMD
which is not mandatory.
In fact, Jain et al.~\cite{JJMPS88} have given an alternative
derivation of the $\pi\rho\omega$ coupling arguing that the consideration
of electromagnetic processes is not theoretically reliable for extracting
$g_{\pi\rho\omega}$.
In that derivation, the sign of $g_{\pi\rho\omega}^{}$ is undetermined.

It is, therefore, clear that the sign of $g_{\pi\rho\omega}^{}$ still remains
to be determined and that the existing data on reaction processes such as
$\pi^- p \to \omega n$ and $pp\to pp\pi^0$ do not impose sufficiently
stringent constraints for determining this sign.
In this paper, we show that $\omega$-meson production in $NN$ collisions,
and specifically the reaction $pp \to pp\omega$, is more suited for the 
determination of the relative sign of $g_{\pi\rho\omega}^{}$ and that the
recent data from the COSY-TOF~\cite{COSY01,COSY-05c} and
COSY-ANKE~\cite{COSY-ANKE-04,COSY-ANKE-06} collaborations in conjunction
with the earlier data \cite{DISTO-98,HBBC99} strongly favor a negative
$g_{\pi\rho\omega}^{}$, in contrast to a positive $g_{\pi\rho\omega}^{}$
used in many calculations of pertinent hadronic reactions.

We follow the Distorted Wave Born Approximation approach employed in 
Refs.~\cite{NSHHS98,NDHHS99,TN03a} in order to describe $\omega$-meson 
production in $NN$ collisions.
The total amplitude $M^\mu$ is written as
\begin{equation}
M^\mu = (1 + T_f G_f)J^\mu(1 + G_i T_i) \ ,
\label{ampl}
\end{equation}
where $T_\lambda$ stands for the $NN$ initial and final state interactions
as $\lambda = i$ and $f$, respectively.
$G_\lambda$ denotes the corresponding two-body $NN$ propagator and
$J^\mu$ denotes the $\omega$-meson production current.
%
\begin{figure}[b!]\centering
\includegraphics[width=0.45\textwidth,angle=0,clip]{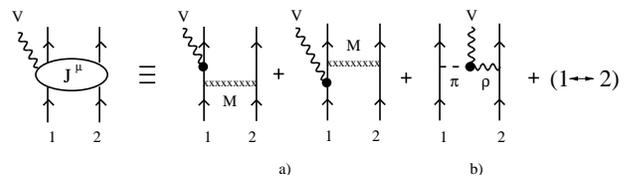}
\caption{\label{fig:curr}%
The basic vector-meson production current $J^\mu$. (a) corresponds to the 
nucleonic current ($M=\sigma, a_0, \eta, \pi, \omega, \rho$), 
while (b) is the mesonic current. $V$ stands for the vector-meson $\omega$.}
\end{figure}
%
%
\begin{figure*}[t!]
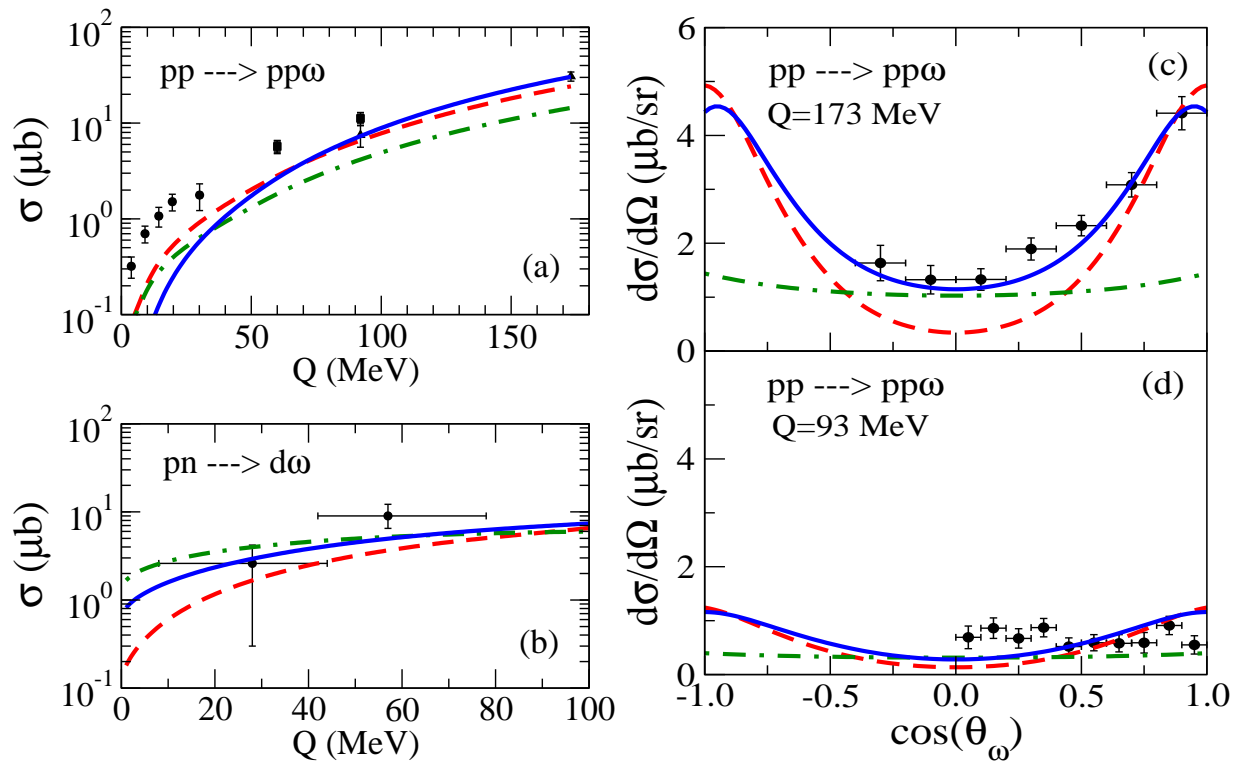
\centering
\includegraphics[width=0.45\textwidth,angle=0,clip]{fig3ab.eps}
\includegraphics[height=0.422\textheight,width=0.45\textwidth,angle=0,clip]
{fig3cd.eps}
\caption{\label{fig:destr}%
Left panel: total cross sections for $pp \to pp\omega$ 
(a), and for $pn \to d\omega$ (b) as a function of the excess energy $Q$. 
Right panel: $\omega$ angular distributions in the overall
center-of-mass frame for $pp \to pp\omega$ at $Q=173$ (c) and 93 MeV (d). 
All results are obtained with $g_{\pi\rho\omega}^{}=+10$ in 
Eq.~(\ref{pirhoomega}).
The dashed curves correspond to the nucleonic and dash-dotted ones to the 
mesonic currents. The solid curves correspond to sum of them, i.e., the total 
current. Data are from
Refs.~\cite{COSY01,COSY-05c,COSY-ANKE-04,COSY-ANKE-06,DISTO-98,HBBC99}.}   
\end{figure*}
%
%
The $NN$ final state interaction (FSI), which is known to introduce
strong energy dependence in this reaction near threshold, is treated exactly,
while the initial state interaction (ISI) is considered within the on-shell
approximation.
We refer to Ref.~\cite{HN99} for the discussion on the validity of the latter
approximation.
Also, the effect of the finite $\omega$ meson width is taken into account in
the present work.
This effect is well known to enhance the cross sections near the threshold
energies \cite{HBBC99,NDHHS99,TN03a}. 
When the final $NN$ state is a bound state, i.e., a deuteron, the reaction
amplitude $M^\mu$ is calculated using the corresponding deuteron wave
function as has been done in Ref.~\cite{NHS00}.  
We consider the nucleonic and mesonic currents as depicted in 
Fig.~\ref{fig:curr} \cite{TN03a}.
The relevant $\omega$-production vertices are given by the Lagrangian
density (\ref{pirhoomega}) for the mesonic current and by 
\begin{equation}
{\cal L}_{NN\omega} = - g_{NN\omega}^{}\bar\Psi \left\{ \left(\gamma_\mu - 
\frac{\kappa_\omega^{}}{2m_N}\sigma_{\mu\nu}\partial^\nu\right)
\omega^\mu\right\}\Psi \ 
\label{NNomega}
\end{equation}
for the nucleonic current.
In the above equation, $\Psi$ stands for the nucleon field and $m_N^{}$
for the nucleon mass.
$g_{NN\omega}^{}$ denotes the vector coupling constant and
$\kappa_\omega^{}$ is the ratio of the tensor to vector coupling constants. 
As far as the relative sign between the nucleonic and mesonic currents
is concerned, we also need to specify the $NN\pi$ and $NN\rho$ vertices 
entering in the mesonic current.
They are obtained from the Lagrangian densities 
\begin{eqnarray}
{\cal L}_{NN\rho} & = & - g_{NN\rho}^{}\bar\Psi \left\{ \left(\gamma_\mu - 
\frac{\kappa_\rho^{}}{2m_N}\sigma_{\mu\nu}\partial^\nu\right)
\vec\tau\cdot\vec\rho^\mu\right\}\Psi \ ,
\nonumber \\
{\cal L}_{NN\pi} & = & - \frac{g_{NN\pi}^{}}{2m_N^{}}
\bar\Psi \gamma_5\gamma_\mu\vec\tau\cdot\left(\partial^\mu\vec\pi \right)
\Psi \ , 
\label{NNpi-rho}
\end{eqnarray}
where $g_{NN\rho}^{}=3.36$, $\kappa_\rho^{}=6.1$ and $g_{NN\pi}^{}=13.45$.

Figure~\ref{fig:destr} shows our results for the total cross sections for
$pp\to pp\omega$ and for $pn \to d\omega$, as well as the $\omega$ angular
distribution in $pp\to pp\omega$.
The free parameters of our model
---
the cutoff masses of the form factors at the $NN\omega$ and
$\pi\rho\omega$ vertices, $\Lambda_N$ and $\Lambda_M$, respectively,
(see Ref.~\cite{TN03a} for details) and the ratio of the tensor to
vector coupling constant, $\kappa_\omega^{}$, in Eq.~(\ref{NNomega}) 
---
were adjusted to reproduce the total and differential cross section data for
$pp\to pp\omega$ at $Q=173$ MeV.
In doing so, we include the $NN$ ISI and FSI. 
The resulting parameter values are $\Lambda_N=1200$ MeV, $\Lambda_M=1120$ MeV
and $\kappa_\omega^{}=-2$.
Here, a somewhat large value of $|\kappa_\omega^{}|$ is required to reproduce
the shape of the measured angular distribution at $Q=173$ MeV.
It can be brought down to a more reasonable value of
$\kappa_\omega^{} \sim -0.5$ once the resonance currents are considered as
pointed out in Ref.~\cite{TN03a}.
In this context we mention that the value of $\kappa_\omega$ influences 
the energy dependence of the total cross section too.
In fact, for vanishing $\kappa_\omega$ the predicted energy dependence 
of the total cross section for $pp \to pp\omega$ is considerably better than
that shown in Fig.~\ref{fig:destr}(a).
However, with such a value, a satisfactory description of the measured
angular distributions is no longer possible.

The result for the $pn \to d\omega$ reaction (Fig.~\ref{fig:destr}(b))
was obtained with the same model parameters as used for $pp \to pp\omega$. 
In addition, we assumed that the ISI causes the same reduction of the total
cross section as it does in the $pp$-induced reaction.
We should mention, however, that the validity of this assumption is debatable.
Indeed, the experimental information on the $pn$ interaction for
laboratory energies relevant for $\omega$ production whose threshold-energy
is at around $T_{\rm lab}=1.89$ GeV is rather poor.
Specifically, for $T=0$ there is no $NN$ phase-shift analysis available
for energies above $T_{\rm lab}=1.3$ GeV.
At the latter energy, the reduction factor due to the ISI, evaluated according
to the prescription given in Ref.~\cite{HN99}, is about $0.3$.
On the other hand, the corresponding reduction factor for
$pp \to pp\omega$ ($T=1$) at $T_{\rm lab}=1.89$ GeV is about 0.45. 
These values may provide a very rough idea on the uncertainty associated with 
the present procedure to account for the $pn$ ISI.
However, in view of the large error bars of the $pn \to d\omega$ data the
mentioned ambiguities in the ISI are not really significant and one can
certainly say that our model results for the $pn$ induced reaction are in
line with the experimental data.

The results in Fig.~\ref{fig:destr} were obtained with the values of the
coupling constants $g_{NN\omega}= 9$ and $g_{\pi\rho\omega}=+10$, which lead
to a strong destructive interference between the nucleonic and mesonic current 
contributions%
\footnote{Here, the term destructive or constructive interference will be
employed in the sense of the resulting cross section being smaller or 
larger than the sum of the individual contributions.},
especially at lower excess energies in $pp\to pp\omega$.
The value of $g_{\pi\rho\omega}=+10$ has been extracted from the measured 
$\omega \to \pi^0 \gamma$ radiative decay rate in conjunction with the VMD 
assumption.
The sign of this coupling constant is determined by the sign of the
$\pi\omega\gamma$ coupling constant which, in turn, has been fixed from 
the analysis of pion photoproduction in the 1 GeV energy region \cite{GM93}. 
As a consequence of this interference pattern, the predicted energy dependence 
of the total cross section for $pp\to pp\omega$ is in serious disagreement
with the experimental information.
Specifically, the model calculation (with parameters adjusted to the data
at higher energies) strongly underestimates the data at near-threshold
energies.
We note that other authors who have investigated $\omega$-meson
production in $NN$ collisions \cite{Kaiser99,GKB00} have used the same 
relative sign between the nucleonic and mesonic currents as in 
Refs.~\cite{NSHHS98,NDHHS99,TN03a}.

A possible mechanism to cure the discrepancy observed above is the 
excitation of nucleon resonances.
In fact, in Ref.~\cite{TN03a} nucleon resonance contributions have been
explored in $pp\to pp\omega$; however, these were found to be insufficient
to provide the necessary enhancement of the total cross section near
threshold in order to reproduce the data.
Another possibility to remedy the problem is the (background) $\omega N$ FSI. 
In the case of $\eta$ production in $NN$ collisions, it is generally believed 
that the $\eta N$ FSI is responsible for the experimentally observed 
enhancement of the cross section near threshold by a factor of two or so. 
Only its actual strength is still under debate as reflected by the values of 
the $\eta N$ scattering length one can find in the literature: 
$a_{\eta N} = (0.2 \sim 1.1, 0.26 \sim 0.35)$~fm \cite{SSEHKS01}. 
In comparison, the estimated (spin-averaged) $\omega N$ scattering length is
of the order of $a_{\omega N} = (-0.026 \sim 1.6, 0.20 \sim 0.30)$~fm
\cite{SLMP04,LWF01,KWW99}.
Therefore, one might expect also some effects of the $\omega N$ FSI in
$pp \to pp\omega$.
However, in any case, it is not trivial to come up with new mechanisms
which could enhance the $pp \to pp\omega$ cross section close to threshold
by more than an order of magnitude and, \emph{at the same time\/}, leave the 
cross section in $pn \to d\omega$ more or less unchanged in order to solve the 
observed discrepancies.

\begin{figure*}[t!]
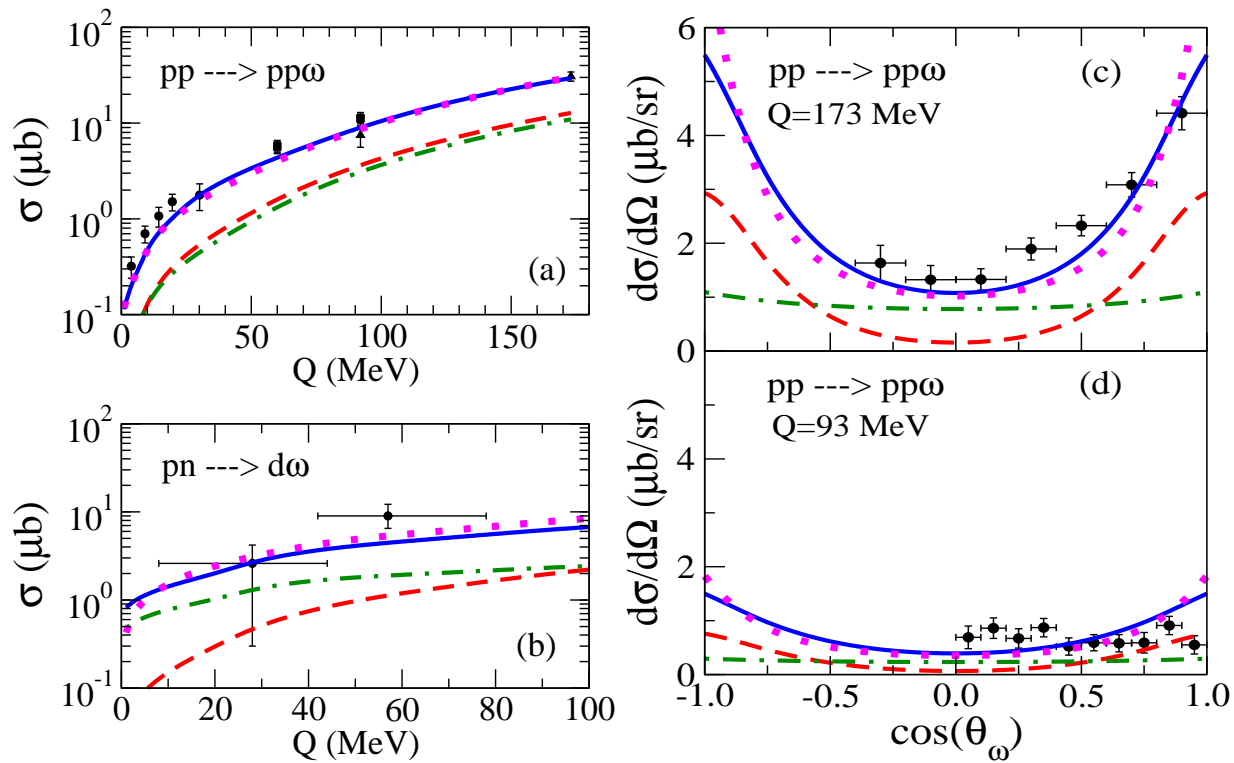
\centering
\includegraphics[width=0.45\textwidth,angle=0,clip]{fig4ab.eps}
\includegraphics[height=0.422\textheight,width=0.45\textwidth,angle=0,clip]%
{fig4cd.eps}
\caption{\label{fig:const}%
Same as Fig.~\ref{fig:destr} with the choice 
$g_{\pi\rho\omega}^{}=-10$.
The dotted curves correspond to the results with $\kappa_\omega=0$.}   
\end{figure*}
%

In the following, we show that the discrepancy discussed above can be largely
eliminated if one changes the sign of the $\pi\rho\omega$ coupling constant
with respect to the $NN\omega$ coupling constant, i.e., if one assumes a 
negative coupling constant $g_{\pi\rho\omega}^{}$ in
Eq.~(\ref{pirhoomega}).
The change of the sign is motivated by the following two observations. 
First, we note that the isospin operator structures in the nucleonic
current are {\bf 1} for the isoscalar meson exchanges ($M = \sigma, \eta, 
\omega$) and $\vec\tau_1\cdot\vec\tau_2$ for the isovector meson 
exchanges ($M = a_0, \pi, \rho$).
The structure of the mesonic current is $\vec\tau_1\cdot\vec\tau_2$, i.e.
the same as that for the isovector meson exchanges in the nucleonic current.
The isospin matrix element is then $1$ for $pp\to pp\omega$
(total isospin $T=1$) and $-3$ for $pn \to d\omega$ ($T=0$) in the mesonic
current and in the part of the nucleonic current involving isovector meson
exchanges, while it is always $1$ in the nucleonic current involving the
isoscalar meson exchanges. 
Second, the qualitative features of the calculated total cross sections close
to the threshold displayed in Figs.~\ref{fig:destr}(a,b) can be understood 
easily if one assumes that the reaction amplitudes (without the isospin factor)
due to the isoscalar (isovector) meson exchanges in the nucleonic current,
$\alpha$ ($\beta$), and due to the mesonic current, $\gamma$, are related by
$\alpha \sim -2c/3$, $\beta \sim -c/3$ and $\gamma \sim c$, where $c$ is a
complex number.
We then have    
\begin{eqnarray} 
M_{pp\omega} & = & \alpha + \beta + \gamma \sim 0 \ , \nonumber \\
M_{d\omega}  & = & \alpha - 3\beta - 3\gamma \sim - \frac{8}{3}c \ ,
\label{destr_ampl}
\end{eqnarray}
for the $pp \to pp\omega$ and $pn \to d\omega$ total reaction amplitudes, 
$M_{pp\omega}$ and $M_{d\omega}$, respectively.
Note, in particular, that the nucleonic and mesonic current contributions are
practically the same in the $pp$-induced reaction, while in the $pn$-induced
reaction, the mesonic current dominates over the nucleonic current.
Moreover, in the latter reaction, the total current contribution is smaller
than the mesonic and larger than the nucleonic current contributions.    
If we now change the sign of the mesonic current, we have 
\begin{eqnarray}
M_{pp\omega} & = & \alpha + \beta - \gamma \sim  - 2c \ , \nonumber \\
M_{d\omega}  & = & \alpha - 3\beta + 3\gamma \sim \frac{10}{3}c \ ,
\label{constr_ampl}
\end{eqnarray}
for the corresponding total reaction amplitudes.
This seems precisely what is required to reproduce the measured cross
sections in Figs.~\ref{fig:destr}(a,b): an enhancement of the total cross
section close to threshold in $pp \to pp\omega$ and practically no
change in $pn \to d\omega$.

In Fig.~\ref{fig:const} our results with $g_{NN\omega}^{}=9$ and 
$g_{\pi\rho\omega}^{}=-10$ are displayed.
We follow the same strategy for adjusting the model parameters as described 
above for the case of a positive $\pi\rho\omega$ coupling constant.
Now the resulting parameter values are $\Lambda_N=1100$ MeV,
$\Lambda_M=1000$ MeV, and $\kappa_\omega^{}=-2$. 
As can be seen, in contrast to the predictions in Fig.~\ref{fig:destr} for 
$g_{\pi\rho\omega}^{}=+10$, with the negative sign the total cross section 
for $pp \to pp\omega$ is in much better agreement with the data.
Specifically, the underestimation of the cross section near the threshold by
more than an order of magnitude (cf. Fig.~\ref{fig:destr}) is now strongly
reduced to only a factor of about 1.5.
At the same time the reaction $pn \to d\omega$ is again in qualitative
agreement with the data (Fig.~\ref{fig:const}b).
The dotted curves in Fig.~\ref{fig:const} correspond to results
with $\Lambda_N=990$ MeV, $\Lambda_M=950$ MeV, and $\kappa_\omega^{}=0$.
Contrary to the case with positive $g_{\pi\rho\omega}^{}$, where a value 
of about $\kappa_\omega^{}=-2$ is required to reproduce the shape of the
measured $\omega$ angular distribution, now, with a negative value of
$g_{\pi\rho\omega}^{}$, the existing data can be described rather reasonably
even with a vanishing $\kappa_\omega^{}$.
Note that $\kappa_\omega^{}\approx 0$ is more in line with the values
employed in other reactions such as $NN$ scattering.

As discussed above, it is very reasonable to expect that nucleon resonances
and (background) $\omega N$ FSI would bring the prediction in
Fig.~\ref{fig:const} in even better agreement with the data once they are
taken into account together. 
Efforts to include them consistently with other more basic reactions such as 
$\pi N \to \omega N$ and $\gamma N \to \omega N$ are currently in progress.

In this context we want to emphasize that our results with the negative 
coupling constant $g_{\pi\rho\omega}^{}$ are actually in line with 
the recent calculations presented in Ref.~\cite{KK04}, once we neglect the
effect of the $\omega$ meson width and use a constant reduction factor to
simulate the $NN$ ISI instead of its explicit inclusion.
As mentioned before, taking into account the $\omega$ meson width enhances
the cross sections near threshold \cite{HBBC99,NDHHS99,TN03a}.
On the other hand, the explicit inclusion of the ISI in our investigation
introduces a significant energy dependence over the energy range considered.
If the model parameters are adjusted to reproduce the data at higher energies,
as in our calculation, then one observes an underestimation of the
total cross section close to threshold, (cf. Fig.~\ref{fig:const}(a)).
Note that neither finite width effects nor the explicit inclusion of the
$NN$ ISI were considered in Ref.~\cite{KK04}.

In summary, we have shown that the relative sign of the $\pi\rho\omega$
coupling constant in Eq.~(\ref{pirhoomega}) may be most sensitively
determined from $\omega$ meson production in $NN$ collisions, due to the
distinct isospin structures of the nucleonic and mesonic currents.
Other hadronic reactions such as $\pi N \to \omega N$, where 
the isospin structure is the same for both the (corresponding)
nucleonic and mesonic currents, are certainly much less suited for fixing
the sign of the $\pi\rho\omega$ coupling constant.

Indeed, according to our results the existing data for $pp \to pp\omega$
strongly favor a negative sign of the coupling constant
$g_{\pi\rho\omega}^{}$ in Eq.~(\ref{pirhoomega}), in contrast to the positive
sign as inferred from studies of the pion photoproduction reaction in
conjunction with the VMD assumption.
With regard to the latter, one should keep in mind that in the most
investigated $\Delta$-excitation region the fits to the pion
photoproduction data are not particularly sensitive to the $t$-channel 
$\omega$-exchange mechanism which provides only about 5\% or less of the
$\pi^0$ photoproduction total cross section at $E_\gamma = 300$ MeV within
both the Sato-Lee \cite{SL96} and the Haberzettl-Nakayama-Krewald \cite{HNK06}
models.
Changing the sign of the $\omega$-exchange current will not influence the 
main results of those studies too much, although it may help to remove
some remaining discrepancies with the data.
However, the $\omega$-exchange mechanism becomes very large in the 
higher-mass nucleon resonance region.
For example, changing the sign in the $\omega$-exchange contribution 
would increase the total $\pi^0$ photoproduction cross section by a
factor of $\sim 1.5$ at $E_\gamma = 800$ MeV in a calculation including only
the Born term and $\omega$-exchange current and using the parameters of
Ref.~\cite{HNK06}.
Hence, the use of the correct sign of $\omega$-exchange could be crucial in
defining the non-resonant amplitudes for extracting the higher-mass nucleon
resonance parameters from the data.
Therefore, we expect that the sign of $g_{\pi\rho\omega}^{}$ as fixed in
this work will have an impact on the corresponding nucleon resonance
parameters too.

Finally, we mention that the present result is, admittedly, model dependent.
Indeed, our calculations are based on a model containing a few 
free parameters that have been adjusted to reproduce the data.  
However, we believe that our conclusion on the sign of
$g_{\pi\rho\omega}^{}$ is rather robust, given the relatively strong effect
of the interference between the nucleonic and mesonic currents on the energy
dependence of the total cross sections.
Although there are other possible production mechanisms ignored in the
present calculation, their effects would not be strong enough to change
our conclusion, as can be inferred from our knowledge
in the study of meson production reactions in $NN$ collisions.

\acknowledgments

We thank Toru Sato, Horst Lenske and Norbert Kaiser for checking the
relative signs of the $\omega\pi\gamma$ and $\pi\rho\omega$ coupling
constants used in their calculations. 
This work was supported in part by the FFE Grant No. 41445282 (COSY-58)
and by the U.S. Department of Energy, Office of Nuclear Physics, under
Contract No. W-31-109-ENG-38.

\end{document}